\documentclass[12pt]{article}
\usepackage{amsmath,amsbsy,amsfonts,amssymb,graphics,epsfig,float,bm,verbatim,subfigure}

\def\fakebold#1{\relax\ifvmode\leavevmode\fi%
                                                                                                        
\ifmmode%
\setbox0=\hbox{$#1$}%
\else%
\setbox0=\hbox{#1}%
\fi%
\kern-.02em\copy0 \kern-\wd0%
\kern .04em\copy0 \kern-\wd0%
\kern-.0125em\raise.02em\box0%
}%

\renewcommand{\geq}{\geqslant}

\renewcommand{\vec}[1]{\boldsymbol{#1}}
\newcommand{\ii}{\textrm{i}}
\newcommand{\ee}{\textrm{e}}
\newcommand{\I}{\textrm{I}}
\newcommand{\K}{\textrm{K}}

\numberwithin{equation}{section}

\begin{document}

\title{Nonlinear Euler buckling}
\author{
  A. Goriely,  R. Vandiver, M. Destrade}
\date{2008}
\maketitle

\begin{abstract}
The buckling of hyperelastic incompressible cylindrical shells 
of arbitrary length and thickness under axial load is considered within 
the framework of nonlinear elasticity. 
Analytical and numerical methods 
for bifurcation are obtained using the Stroh formalism and the exact solution 
of Wilkes [\emph{Q. J. Mech. Appl. Math.} 8:88--100, 1955] for the linearized problem. 
The main focus of this paper is the range of validity of the Euler 
buckling formula and its first nonlinear corrections 
that are obtained for third-order elasticity.

\end{abstract}

\section{Introduction}

\begin{figure}[!ht]
 \centering 
\includegraphics[width=0.4\textwidth]{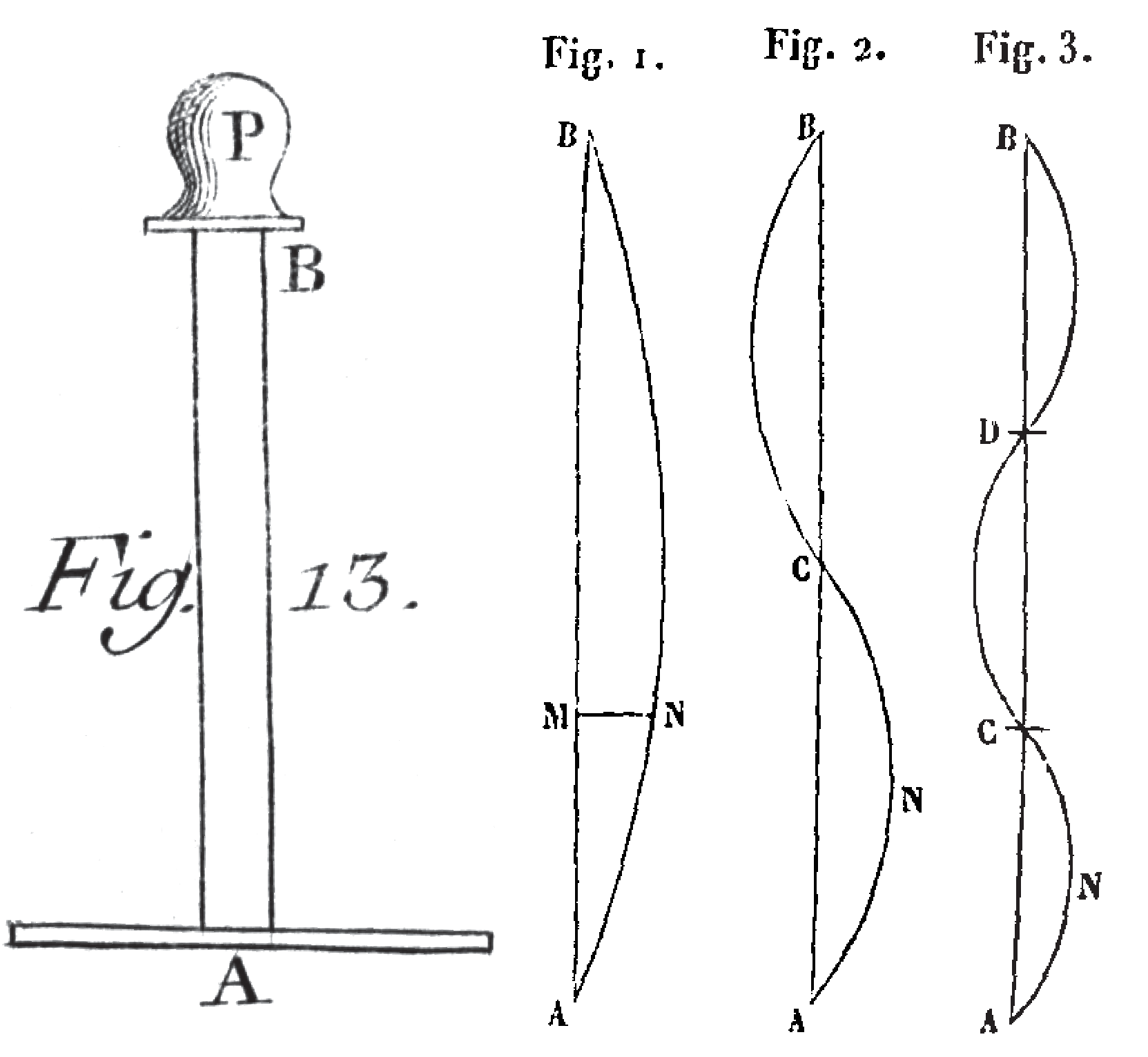}
 \caption{Euler problem: Left: illustrations from Euler manuscript \cite{eu44}. Right: Lagrange solutions \cite{la70}, mode 1, 2, and 3.}
  \label{figeuler}
\end{figure}

Under a large enough axial load an elastic beam will buckle. This phenomenon known as \textit{elastic buckling}  or \textit{Euler buckling} is one of the most celebrated instabilities of classical elasticity. The critical load for buckling was first derived by Euler in 1744 \cite{eu44,olelbr33,eu59} and further refined for higher modes by Lagrange in 1770 \cite{la70,ti83}. Both authors reached their conclusion on the basis of simple beam equations first derived by Bernoulli \cite{to93} (See Fig.~\ref{figeuler}). Since then, Euler buckling has played a central role in the stability and mechanical properties of slender structures from nano- to macro- structures in physics, engineering, biochemistry, and biology \cite{tige61,ni92}. 
Explicitly, the critical compressive axial load $N$ that will lead to a buckling instability of a hinged-hinged isotropic homogeneous beam of length $L$ is 
\begin{equation}\label{euler}
N_{\textrm{Euler}}= \dfrac{\pi ^2 E I}{L^2}, 
\end{equation}
where ``$\pi$ is the circumference of a circle whose diameter is one" \cite{eu59}, $E$ is Young's Modulus, and $I$ is the second moment of area which, in the case of a cylindrical shell of inner radius $A$ and outer radius $B$, is $I=\pi(B^4-A^4)/4$.

There are many different ways to obtain this critical value and infinite variations on the theme. If the beam is seen as a long slender structure, the \textit{one-dimensional} theory of beams, elastica, or Kirchhoff rods, can be used successfully to capture the instability, either by bifurcation analysis, energy argument \cite{tige61}, or directly from the exact solution, which in the case of rods can be written in terms of elliptic integrals \cite{nigo99}. The one-dimensional theory  can be used with a variety of boundary conditions, it is particularly easy to explain and generalize, and it can be used for large geometric deflections of the axis \cite{an95}. However, since material cross sections initially perpendicular to the axis remain undeformed and perpendicular to the tangent vector, no information on the elastic deformation around the central curve can be obtained. In particular, other modes of instability such as barreling cannot be obtained. Here, by \textit{barreling}, we refer to axisymmetric deformation modes of a cylinder or a cylindrical shell. These modes will typically occur for sufficiently short structure.

The \textit{two-dimensional} theory of shells can be used when the thickness of the cylindrical shell is small enough. Then the stability analysis of shell equations such as the  Donnell-von K\'{a}rm\'{a}n equations leads to detailed  information on symmetric instability modes, their localization and  selection \cite{hulope03}. However, the theory cannot be directly applied to obtain information on the buckling instability (asymmetric buckling mode).

The \textit{three-dimensional} theory of nonlinear elasticity provides, in principle, a complete and exact description of the motion of each material point of a body under loads. However, due to the mathematical complexity  of the governing equations,  most problems cannot be explicitly solved. In the case of long slender structures under loads, the buckling instability and its asymptotic limit to Euler criterion can be captured by assuming that the object is either a rectangular beam \cite{bi62,le68,no69} or a cylindrical shell under axial load. Then, using the theory of incremental deformations around a large deformation stressed state, the buckling instability can be recovered by a bifurcation argument, usually referred to, in the nonlinear elasticity theory, as \textit{small-on-large}, or \emph{incremental}, deformations. In comparison to the one- and two-dimensional theories, this computation is rather cumbersome as it is based on non-trivial tensorial calculations, but it contains much information about the instability and the unstable modes selected in the bifurcation process. 

 Here, we are concerned with the case of a cylindrical shell under axial load. This problem was first addressed in the framework of nonlinear elasticity in a remarkable 1955 paper by Wilkes \cite{wi55} who showed that the linearized system around a finite axial strain can be solved exactly in terms of Bessel functions. While Wilkes only analyzed the first axisymmetric mode ($n=0$, see below), he noted in his conclusion that the asymmetric mode ($n=1$) corresponds to the Euler strut and doing so, opened the door to further investigation by Fosdick and Shield  \cite{fosh63} who recovered Euler's criterion asymptotically from Wilkes's solution. These initial results constitute the basis for much of the modern theory of elastic stability of cylinders within the framework of three-dimensional nonlinear elasticity\cite{sisp84, duensp93,pabe97,bige01,doha06}.  The experimental verification of Euler's criterion was considered by Southwell \cite{so32} and by Beatty and Hook \cite{beho68}. 

The purpose of this paper is threefold. First, we revisit the problem of the stability of an incompressible cylindrical shell under axial load using the Stroh formalism \cite{st62} and, based on Wilkes solution, we derive a new and compact formulation of the bifurcation criterion that can be used efficiently for numerical approximation of the bifurcation curves for all modes. Second, we use this formulation to obtain  nonlinear corrections of Euler's criterion for arbitrary shell thickness and third-order elasticity. Third, we consider the problem of determining the critical aspect ratio where there is a transition between buckling and barreling.

\section{Large deformation}


We consider a hyperelastic homogeneous incompressible cylindrical tube of initial inner radius $A$, outer
radius $B$, and length $L$, subjected to a uniaxial constant strain
$\lambda_3$, and deformed into a shorter tube with current dimensions
$a$, $b$, and $l$. 
The deformation bringing a point at ($R$, $\Theta$, $Z$),  in cylindrical coordinates in the initial
configuration, to ($r$, $\theta$, $z$) in the current configuration is 
\begin{equation}
 r = \lambda_1 R, \qquad \theta = \Theta, \qquad z = \lambda_3 Z,
 \label{deformation}
\end{equation}
where $\lambda_1 = a/A = b/B$ and $\lambda_3 = l/L$.
The physical components of the corresponding deformation gradient
$\vec{F}$ are
\begin{equation}
 [\vec{F}] = \text{diag }(\lambda_1, \lambda_1, \lambda_3),
\end{equation}
showing that the principal stretches are the constants $\lambda_1$,
$\lambda_2 = \lambda_1$, $\lambda_3$; and that the pre-strain is
homogeneous. 
Because of incompressibility, $\text{det }\vec{F} = 1$, so that
\begin{equation} \label{pre-strain}
 \lambda_1 =  \lambda_3^{-1/2}.
\end{equation}
The principal Cauchy stresses required to maintain the pre-strain are
\cite{og84} 
\begin{equation}
 \sigma_i = -p + \lambda_i \dfrac{\partial W}{\partial \lambda_i}, \qquad
  (i = 1, 2, 3),
\end{equation}
(no sum) where $p$ is a Lagrange multiplier introduced by the internal
constraint of incompressibility and $W$ is the strain energy
density (a symmetric function of the principal stretches).
In our case, $\sigma_2 = \sigma_1$ because $\lambda_2 = \lambda_1$. 
Also, $\sigma_1 = 0$ because the inner and outer faces of the tube are
free of traction. It follows that
\begin{equation}\label{s3}
 p = \lambda_1 W_1, \qquad \sigma_3 = \lambda_3 W_3 -
\lambda_1 W_1,
\end{equation}
where $W_i \equiv \partial W / \partial \lambda_i$, and we conclude that
the principal Cauchy stresses are constant.

                                              
\section{Instability}

 
To perform a bifurcation analysis, we take the view that the existence of 
small deformation solutions in the neighborhood of the large pre-strain
signals the onset of instability \cite{bi65}. 

                                              
\subsection{Governing equations}

 
The incremental equations of equilibrium and incompressibility read 
\cite{og84}
\begin{equation} \label{governing}
\text{div } \vec{s} = \vec{0}, \qquad \text{div }\vec{u} = 0,
\end{equation}
where $\vec{s}$ is the incremental nominal stress tensor and $\vec{u}$
is the infinitesimal mechanical displacement. 
They are linked by 
\begin{equation}
 \vec{s} = \vec{\mathcal{A}}_\mathbf{0} \text{ grad} (\vec{u})
  + p \text{ grad} (\vec{u}) - \dot{p} \vec{I},
\end{equation}
where $\dot{p}$ is the increment in the Lagrange multiplier $p$ and 
$\vec{\mathcal{A}_\mathbf{0}}$ is the fourth-order tensor of
instantaneous elastic moduli.
This tensor is similar to the stiffness tensor of linear anisotropic
elasticity, with the differences that it possesses only the minor
symmetries, not the major ones, and that it reflects strain-induced
anisotropy instead of intrinsic anisotropy.
Its explicit non-zero components in a coordinate system aligned with
the principal axes are \cite{og84}:
\begin{align} \label{A0_principal}
& \mathcal{A}_{0iijj} \lambda_i \lambda_j W_{ij}, &&
\nonumber \\
& \mathcal{A}_{0ijij}   (\lambda_i W_i-\lambda_j
W_j)\lambda_i^2/(\lambda_i^2-\lambda_j^2) && \text{if} \quad
i \neq j, \;\lambda_i \neq \lambda_j, 
\nonumber \\
& \mathcal{A}_{0ijij}   (\mathcal{A}_{0iiii} - \mathcal{A}_{0iijj} + \lambda_i W_i)/2
&& \text{if} \quad i \neq j,\;\lambda_i = \lambda_j,
\nonumber \\
& \mathcal{A}_{0ijji}= \mathcal{A}_{0jiij}   \mathcal{A}_{0ijij} - \lambda_i W_i, &&
\end{align}
(no sums)
where $W_{i j} \equiv \partial^2 W / (\partial \lambda_i
\partial \lambda_j)$.
Note that some of these components are not independent because here
$\lambda_1 = \lambda_2$. In particular we have
\begin{align}
&      \mathcal{A}_{02121}= \mathcal{A}_{01212}, 
\qquad \mathcal{A}_{02323}= \mathcal{A}_{01313},
\qquad \mathcal{A}_{02222}= \mathcal{A}_{01111},
\\ \notag
&      \mathcal{A}_{02233}= \mathcal{A}_{01133},
\qquad \mathcal{A}_{02332}= \mathcal{A}_{01331}, 
\qquad \mathcal{A}_{03232}= \mathcal{A}_{03131},
\\ \notag
&  \mathcal{A}_{01221} + \mathcal{A}_{01212}  \mathcal{A}_{01111} - \mathcal{A}_{01122} = 
 2 \mathcal{A}_{01212} + \mathcal{A}_{01331} - \mathcal{A}_{01313}.
\end{align}

                                              
\subsection{Solutions}

 
We look for solutions which are periodic along the
circumferential and axial directions, and have yet unknown variations
through the thickness of the tube, so that our ansatz is
\begin{multline}
 \label{soln}
\{u, v, w, \dot{p}, s_{rr}, s_{r\theta},s_{rz} \} =  \\
 \{ U(r), V(r), W(r), P(r), S_{rr}(r), S_{r \theta}(r), S_{r z}(r) \}
\ee^{\ii(n\theta + k z)}, 
\end{multline}
where $n = 0, 1, 2, \ldots$ is the \textit{circumferential number},
$k$ is the \textit{axial ``wave''-number}, and all upper-case functions are functions of $r$ alone.

The specialization of the governing equations \eqref{governing}  
to this type of solutions has already been conducted in several
articles 
(see for instance \cite{wi55,fosh63,pabe97,ma89,ne99}, and 
Dorfmann and Haugton \cite{doha06} for the compressible counterpart.)
Here we adapt the work of Shuvalov \cite{sh03} on waves in anisotropic 
cylinders to develop a Stroh-like formulation of the problem \cite{st62}.
The central idea is to introduce  a displacement-traction vector,
\begin{equation}
\vec{\eta} \equiv [U, V, W, \ii r S_{rr}, \ii r S_{r \theta}, \ii r S_{rz}]^t,
\end{equation}
so that the incremental equations can be written in the form
\begin{equation} \label{stroh}
\dfrac{\text{d}}{\text{d} r} \vec{\eta}(r) = \dfrac{\ii}{r} \vec{G}(r) \vec{\eta}(r),
\end{equation}
where $\vec{G}$ is a $6 \times 6$ matrix, with the block structure
\begin{equation}
\vec{G} = \begin{bmatrix} \vec{G}_1 & \vec{G}_2 \\[2pt] 
                    \vec{G}_3 & \vec{G}^+_1 \end{bmatrix},
  \qquad \vec{G}_2 = \vec{G}_2^+,    \qquad \vec{G}_3 = \vec{G}_3^+.
\end{equation}
Here the superscript `$+$' denotes the Hermitian adjoint (transpose of
the complex conjugate) and 
$\vec{G}_1$, $\vec{G}_2$, $\vec{G}_3$ are the $3 \times 3$ matrices
\begin{equation}
   \begin{bmatrix} 
 \ii & -n & - k r
  \\[6pt]
 -n  &  - \ii & 0 
  \\[10pt]
 - k r & 0 & 0 
            \end{bmatrix},
\qquad
  \begin{bmatrix} 
 0 & 0 & 0
  \\
 0  &  -  \dfrac{1}{\mathcal{A}_{01212}}  & 0 
  \\
  0 & 0 & - \dfrac{1}{\mathcal{A}_{01313}} 
            \end{bmatrix},
\qquad 
\begin{bmatrix} 
 \kappa_{11} & \ii \kappa_{12} & \kappa_{13}
  \\
 - \ii \kappa_{12} & \kappa_{22} & \kappa_{23}
 \\
  - \ii \kappa_{13} & \kappa_{23} & \kappa_{33}
       \end{bmatrix},
\end{equation}
respectively, with 
\begin{align}
& \kappa_{11} = 4\mathcal{A}_{01212} + (\mathcal{A}_{03131} - \mathcal{A}_{01313}) k^2 r^2,
 \qquad
 \kappa_{23} = n(2 \mathcal{A}_{01212} + \mathcal{A}_{01313}) k r,
 \notag \\[4pt]
& \kappa_{12} = 4 n \mathcal{A}_{01212},
\qquad
 \kappa_{13} = 2 \mathcal{A}_{01212} k r,
\qquad
 \kappa_{22} =  
   4 n^2 \mathcal{A}_{01212} + \mathcal{A}_{03131} k^2 r^2,
\notag \\[4pt]
& \kappa_{33} = n^2 \mathcal{A}_{01313} + 
  (4 \mathcal{A}_{01212} + 2 \mathcal{A}_{01122} - \mathcal{A}_{01111}
   + \mathcal{A}_{03333} - 2 \mathcal{A}_{01133}) k^2 r^2.
\end{align}

As it happens, there exists a set of 6 explicit Bessel-type solutions to these
equations when $n \neq 0$.
This situation is in marked contrast with the corresponding set-up in 
linear anisotropic elastodynamics, where explicit Bessel-type solutions 
exist only for transversely isotropic cylinders with a set of 4 linearly independent 
modes, and do not exist for cylinders of lesser symmetry \cite{mabe01, sh03b}.
As mentioned in the Introduction, the  6 Bessel solutions are presented in the article by Wilkes \cite{wi55}
(for a derivation see Bigoni and Gei \cite{bige01}).

First, denote by $q_1^2$, $q_2^2$ the roots of the following quadratic in 
$q^2$,
\begin{equation} \label{q1q2}
\mathcal{A}_{01313}q^4 - (\mathcal{A}_{01111} + \mathcal{A}_{03333}
 - 2 \mathcal{A}_{01331} - 2\mathcal{A}_{01133})q^2 + \mathcal{A}_{03131}
  = 0.
\end{equation}
Then the roots of this quadratic are $\pm q_1$ and $\pm q_2$,
and it can be checked that the following two vectors are solutions to
\eqref{stroh},
\begin{multline} \label{eta1eta2}
\vec{\eta}^{(1)}, 
\vec{\eta}^{(2)} \left[ \ii \I'_n(q k r), -\dfrac{n}{q k r}\I_n(q k r), - q \I_n(q k r), \right.
 \\
    -\dfrac{k r}{q}(\mathcal{A}_{01313}q^2 + \mathcal{A}_{03131})\I_n(q k r)
     + 2 \mathcal{A}_{01212}\left(\I'_n(q k r) -\dfrac{n^2}{q k r}\I_n(q k r)\right),
 \\
 \left. -2 \ii n  \mathcal{A}_{01212}\left(\I'_n(q k r) -\dfrac{1}{q k r}\I_n(q k r)\right),
 - \ii (1+q^2) k r  \mathcal{A}_{01313} \I'_n(q k r) \right]^t,
\end{multline}
where $q = q_1, q_2$ in turn and $\I_n$ is the modified 
Bessel function of order $n$.
Similarly, we checked that the following vector $\vec{\eta}^{(3)}$, 
\begin{multline} \label{eta3}
\vec{\eta}^{(3)} = 
\left[ \ii \dfrac{n}{k r} \I_n(q_3 k r), - q_3 \I'_n(q_3 k r),
0, \right.
 \\
    - 2 n q_3  \mathcal{A}_{01212} \left(\I'_n(q_3 k r)  
     - \dfrac{1}{q_3 k r}\I_n(q_3 k r)\right),
 \\
 \ii q_3  \mathcal{A}_{01212} 
      \left(2 \I'_n(q_3 k r) - 
          \left(q_3 k r + 2\dfrac{n^2}{q_3 k r}\right) 
  \I_n(q_3 k r)\right),
  \\
 \left. - \ii n \mathcal{A}_{01313} \I_n(q_3 k r) \right]^t,
\end{multline}
is also a solution, where $q_3$ is root to the quadratic
\begin{equation} \label{q3}
\mathcal{A}_{01212}q^2 - \mathcal{A}_{03131}  = 0.
\end{equation}
Finally we also checked that the vectors $\vec{\eta}^{(4)}$, 
$\vec{\eta}^{(5)}$, and $\vec{\eta}^{(6)}$, obtained by
replacing $\I_n$ with the modified Bessel function $\K_n$ in the
expressions above, are solutions too.

Next,  we follow Shuvalov \cite{sh03} and introduce $\vec{\mathcal{N}}(r)$ as a
\emph{fundamental matrix solution} to \eqref{stroh},
\begin{equation}
\vec{\mathcal{N}}(r) = \left[ \vec{\eta}^{(1)}| 
\vec{\eta}^{(2)}| \ldots| \vec{\eta}^{(6)}\right].
\end{equation}
It clearly satisfies 
\begin{equation} 
\dfrac{\text{d}}{\text{d} r} \vec{\mathcal{N}}(r) 
= \dfrac{\ii}{r} \vec{G}(r) \vec{\mathcal{N}}(r).
\end{equation}
Let $\vec{M}(r,a)$ be the \emph{matricant} solution to \eqref{stroh}, that is the matrix such that 
\begin{equation} \label{M}
\vec{\eta}(r) = \vec{M}(r,a) \vec{\eta}(a), \qquad 
\vec{M}(a,a)  = \vec{I}_{(6)}.
\end{equation}
It is obtained from $\vec{\mathcal{N}}(r)$ (or from any other fundamental matrix 
made of linearly independent combinations of the $\vec{\eta}^{(i)}$) by
\begin{equation} \label{M_N}
 \vec{M}(r,a) = \vec{\mathcal{N}}(r) \vec{\mathcal{N}}^{-1}(a),
\end{equation}
and it has the following block structure
\begin{equation} \label{M_i}
 \vec{M}(r,a) =  \begin{bmatrix} \vec{M}_1(r,a) & \vec{M}_2(r,a) \\[2pt] 
                    \vec{M}_3(r,a) & \vec{M}_4(r,a) \end{bmatrix},
\end{equation}
say.

                                              
\subsection{Boundary conditions}


Some boundary conditions must be enforced on the top and bottom 
faces of the tubes. 
Considering that they remain plane ($W = 0$ on $z=0, l$) and free of incremental 
tractions ($S_{r z} = S_{r \theta} = 0$ on $z = 0, l$) leads to 
\begin{equation}
k = \dfrac{ m \pi}{l} = \dfrac{m \pi}{\lambda_3 L},
\end{equation}
where $m = 1, 2, 3, \ldots$ but, since the equations depend only on $k$, we can take $m=1$ without loss of generality.
 
The other boundary conditions are that the inner and outer faces of the
tube remain free of incremental tractions. 
We call  $\vec{S} \equiv 
[S_{r r}, S_{r \theta}, S_{r z}]^t$ the  traction vector, and $\vec{U}
\equiv [U, V, W]^t$ the displacement vector.
We substitute the condition $\vec{S}(a) = \vec{0}$ into \eqref{M} and
\eqref{M_i} to find the following connection,
\begin{equation}
 r\vec{S}(r) =  \vec{z}(r,a) \vec{U}(r),
 \qquad \text{where} \quad 
 \vec{z} \equiv \ii \vec{M}_3 \vec{M}_1^{-1} 
   \end{equation}
is the (Hermitian) $3 \times 3$ \emph{impedance} \cite{sh03}.
Since $\vec{S}(b)=0$, a non-trivial solution only exists if the matrix $\vec{z}(b,a)$ is singular, which implies 
the \emph{bifurcation condition} 
\begin{equation} \label{bifurcation}
 \text{det } \vec{z} (b,a)  - \ii
 \dfrac{\text{det }\vec{M}_3(b,a)}
         {\text{det }\vec{M}_1(b,a)}
	  = 0.
\end{equation}
This is a \emph{real} equation because $\vec{z} = \vec{z}^+$
\cite{sh03}; 
note that the nature (i.e. real or complex \cite{pabe97}, simple or
double \cite{doha06}) of the roots $q_1$, $q_2$, $q_3$ is irrelevant.


\section{The adjugate method}


We are now in a position to use the bifurcation condition (\ref{bifurcation}) to compute explicitly bifurcation curves for each mode $n$. 
We note that the components of $\vec{\mathcal{A}_0}$ depend on the strain energy
density $W$ and on the pre-strain, which by \eqref{pre-strain}
depends only on $\lambda_3$; so do $q_1, q_2, q_3$ by \eqref{q1q2} and
\eqref{q3}. 
According to \eqref{eta1eta2} and \eqref{eta3}, the entries of
$\vec{M}(b,a)$ thus depend (for a given $W$) on $\lambda_3$, $n$,
$k a $, and $k b$ only. 
For a given material ($W$ specified) with a given thickness ($b/a
= B/A$ specified), the bifurcation equation \eqref{bifurcation} gives a
relationship between a measure of the critical pre-stretch: $\lambda_3 \lambda_1^{-2}$, and a measure of the tube slenderness: $k b = 2 \pi m
(b/l) = 2 \pi m \lambda_3^{-3/2}(B/L)$, for a given bifurcation mode
($n$ specified). That is, for a given tube slenderness, what is the axial strain necessary to excite a given mode.

While this bifurcation condition is formally clear, it has not been successfully implemented to compute all bifurcation curves. Indeed, for mode $n>1$, the numerical root finding of $\text{det}(\vec{z})$ becomes unstable (as observed in \cite{doha06} for a similar problem) and, in explicit computations, most authors do not use the exact solution by Wilkes but use a variety of numerical techniques to solve the linear boundary value problems directly (such as the compound matrix method \cite{haor97}, the determinantal method \cite{bego05}, or the Adams-Moulton method \cite{zhluog08}). Note that from a computational perspective, the Stroh formalism is particularly well-suited and well-behaved \cite{bi85,fu05} and if numerical integration was required it would provide an ideal representation of the governing equation.

Rather than integrating the original linear problem numerically, we now show how to use an alternative form of \eqref{bifurcation} to compute all possible bifurcation curves. This method bypasses the need for numerical integration and reduces the problem to a form that is manageable both numerically and symbolically, to study analytically particular asymptotic limits. The main idea is to transform condition \eqref{bifurcation} by factoring non-vanishing factors. We start by realizing that since the fundamental solutions $\{\vec{\eta}^{(i)},i=1\ldots,6\}$ are linearly independent, the matrix $\vec{\mathcal{N}}(r)$ is invertible for all $r\in [a,b]$, which implies that the elements of $\vec{M}(r,a)$ are bounded for $r\in [a,b]$. Therefore $\det(\vec{M}_1(r,a))$ is bounded and $\det\vec{z}=0$ implies $\det(\vec{M}_3(b,a))=0$. Instead of expressing $\det(\vec{M}_3(b,a))$ as the determinant of a $3\times 3$  submatrix of a  matrix obtained as the product of two $6\times 6$ matrices, we first decompose  $\vec{\mathcal{N}}(r)$ as
\begin{equation} \label{N_i}
 \vec{\mathcal{N}}(r) =  \begin{bmatrix} \vec{\mathcal{N}}_1(r) & \vec{\mathcal{N}}_2(r) \\[2pt] 
                    \vec{\mathcal{N}}_3(r) & \vec{\mathcal{N}}_4(r) \end{bmatrix},
\end{equation}
say, where each block is a $3\times 3$ matrix. We also rewrite Eq. \eqref{M_N} as
\begin{equation}\label{M_N2}
 \vec{M}(r,a) \vec{\mathcal{N}}(a) = \vec{\mathcal{N}}(r).
\end{equation}
and write explicitly the two entries $\vec{\mathcal{N}}_3(r)$ and $\vec{\mathcal{N}}_4(r)$, which are
\begin{eqnarray}
&& \vec{\mathcal{M}}_3(r,a) \vec{\mathcal{N}}_1(a)+\vec{\mathcal{M}}_4(r,a) \vec{\mathcal{N}}_3(a)=\vec{\mathcal{N}}_3(r), \\
&& \vec{\mathcal{M}}_3(r,a) \vec{\mathcal{N}}_2(a)+\vec{\mathcal{M}}_4(r,a) \vec{\mathcal{N}}_4(a)=\vec{\mathcal{N}}_4(r),
\end{eqnarray}
which implies
\begin{multline}
\vec{\mathcal{M}}_3(r,a)  = \\
\left[\vec{\mathcal{N}}_3(r) \vec{\mathcal{N}}_3^{-1}(a) 
  - \vec{\mathcal{N}}_4(r) \vec{\mathcal{N}}_4^{-1}(a)\right]
    \left[ \vec{\mathcal{N}}_1(a) \vec{\mathcal{N}}_3^{-1}(a)-\vec{\mathcal{N}}_2(a) \vec{\mathcal{N}}_4^{-1}(a)\right]^{-1}.
\end{multline}
Using again the fact that the entries of $ \vec{\mathcal{N}}$ are bounded,
we have that the bifurcation condition $\det(\vec{\mathcal{M}}_3(b,a))=0$ implies that
\begin{equation}\label{bif2}
\det\vec{Q}(b,a)=0
\end{equation}
where
\begin{equation}\label{bif3}
\vec{Q}(b,a)= \det(\vec{\mathcal{N}}_4(a)) \vec{\mathcal{N}}_3(b) \mathrm{adj}(\vec{\mathcal{N}}_3(a))-
              \det(\vec{\mathcal{N}}_3(a))\vec{\mathcal{N}}_4(b)  \mathrm{adj}(\vec{\mathcal{N}}_4(a)),
\end{equation}
and adj$(\vec{A})$ is the adjugate matrix of $\vec{A}$, that is the transpose of the cofactor matrix (which in the case of an invertible matrix is simply adj$(\vec{A})$=$\det(\vec{A})\vec{A}^{-1}$). This new bifurcation condition is equivalent to the previous one but has many advantages. The matrix $\vec{Q}$ only involves products of $3\times 3$ matrices and is polynomial in the entries of $\vec{\mathcal{N}}$, that is, $\det\vec{Q}(b,a)$ is a polynomial of degree 18 in Bessel functions and has no denominator (hence no small denominator). Both numerically and symbolically, this determinant is well-behaved, even in the limits $a\to 0$,  which corresponds to a solid cylinder, and $n=0$, which corresponds to the first barreling mode (and usually require a special treatment). We will refer to the use of this form of the bifurcation condition as the \textit{adjugate method}.


\subsection{Numerical results}

\begin{figure}[!ht]
 \centering 
\includegraphics[width=0.5\textwidth]{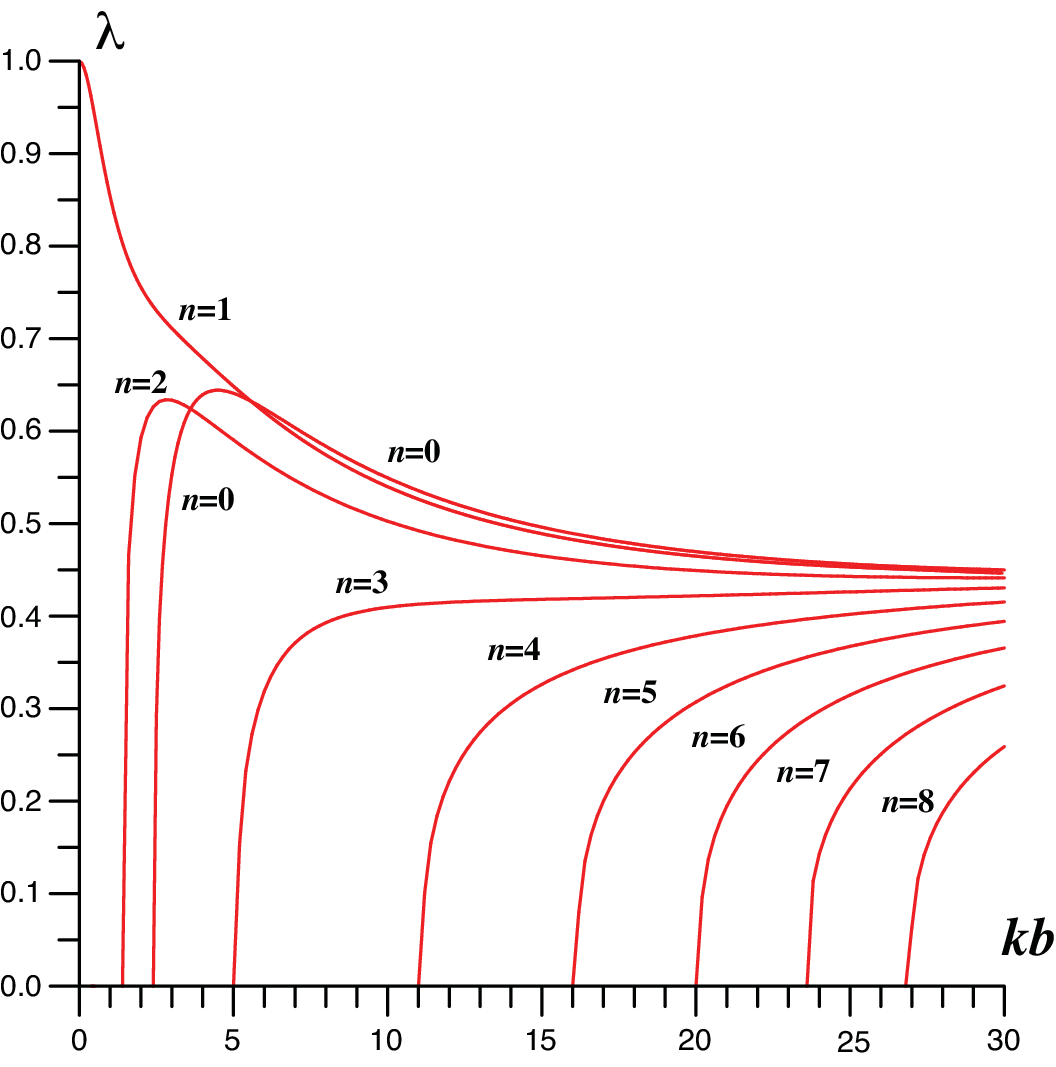}
 \caption{Bifurcation curves (stretch as a function of stubbiness) of an homogeneous neo-Hookean cylindrical tube for modes $n=0$ to $n=9$ with $b/a=B/A=2$ and $C_1=1$.}
  \label{fig2}
\end{figure}

As a first test of the stability of the numerical procedure, we consider a neo-Hookean potential $W= C_1(I_1-3)/2$, for the typical values $C_1=1, B/A=2$ and compute (and plot in Fig. \ref{fig2})  for the ten first modes ($n=0$ to $n=9$), the critical value of $\lambda \equiv \lambda_3$ as a function of the current \emph{stubbiness} $kb=\pi b/l$ (the initial stubbiness is $\nu=B/L=kb \lambda^{3/2}/\pi$). The known classical features of the stability problem for the cylindrical shell are recovered, namely: for slender tubes, the Euler buckling ($n=1$) is dominant and becomes unavoidable as the slenderness decreases; there is a critical slenderness value at which the first barreling mode $n=0$ is the first unstable mode (in a thought experiment where the axial strain would be incrementally increased until the tube becomes unstable); and for very large $kb$, the critical compression ratio tends asymptotically to the value $\lambda=0.444$, which corresponds to surface instability of a compressed half-space \cite{bi62}.

\begin{figure}[!ht]
 \centering 
\includegraphics[width=0.6\textwidth]{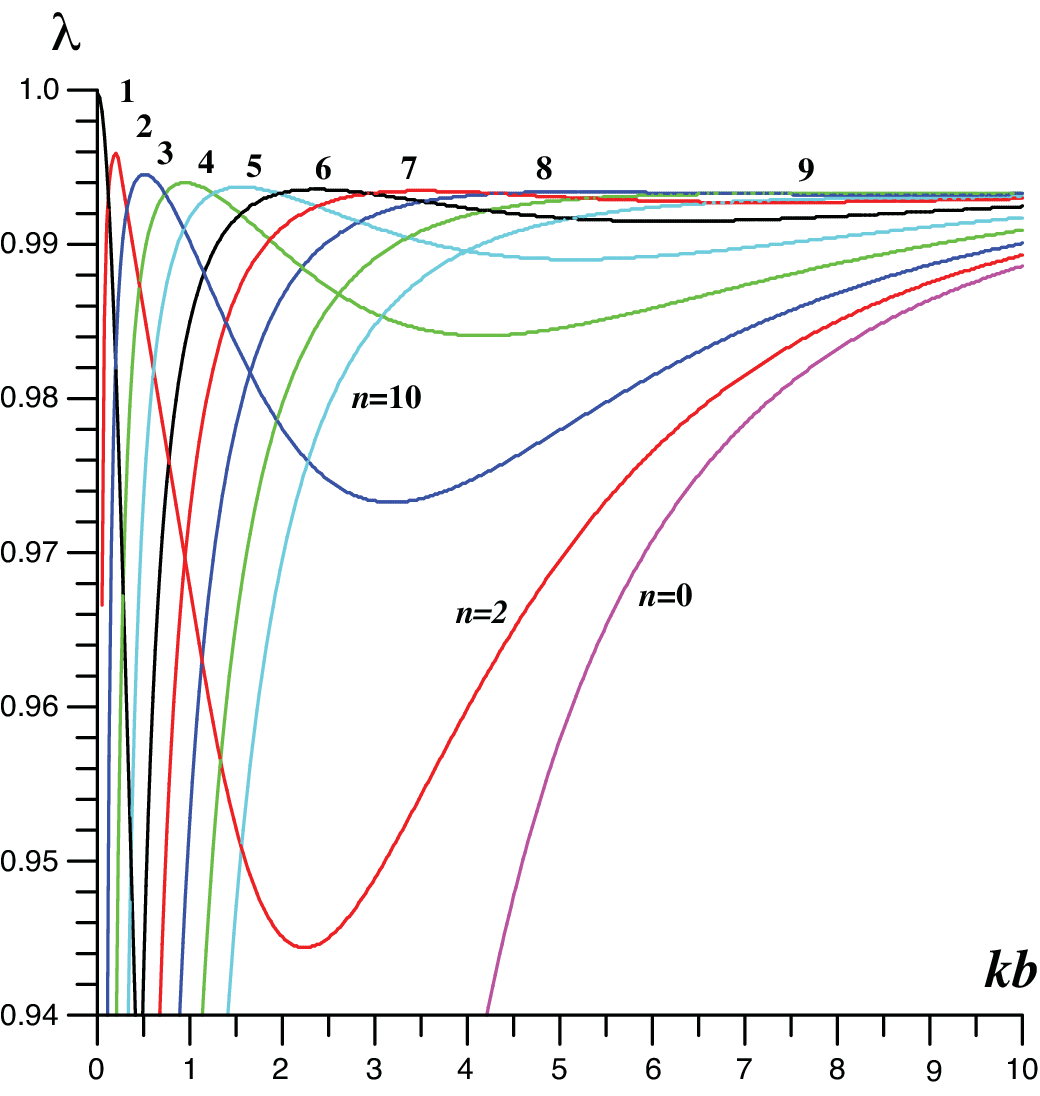}
 \caption{Bifurcation curves (stretch as a function of stubbiness) of an homogeneous neo-Hookean cylindrical tube for modes $n=0$ to $n=10$ with $b/a=B/A=1.01$ and $C_1=1$.}
  \label{fig3}
\end{figure}

For a second test, we consider very thin neo-Hookean tubes with $B/A=1.01$. Here we are interested in the mode selection process. As the stubbiness increases, the buckling mode rapidly ceases to be the first excited mode and is replaced by different barreling modes. From Fig.~\ref{fig3}, it appears clearly that as $kb$ increases, modes $n=1$ to 9 are selected (modes $n=0$ and $n=10$ remain unobservable). There is one particularly interesting feature in these two sets of  bifurcation curves. Depending on both the tube thickness and stubbiness,  the instability mode of a tube transition from buckling to barreling, the material transition from either  the one-dimensional behavior of slender column to the two-dimensional behavior of a thin short tube, or the three-dimensional behavior of a thick short tube. Accordingly we will refer to these particular geometric values where transition occurs as \textit{dimensional transitions} and obtain analytical estimates for them in the next section.


\section{Asymptotic Euler buckling}

\begin{figure}[!ht]
 \centering 
\includegraphics[width=0.8\textwidth]{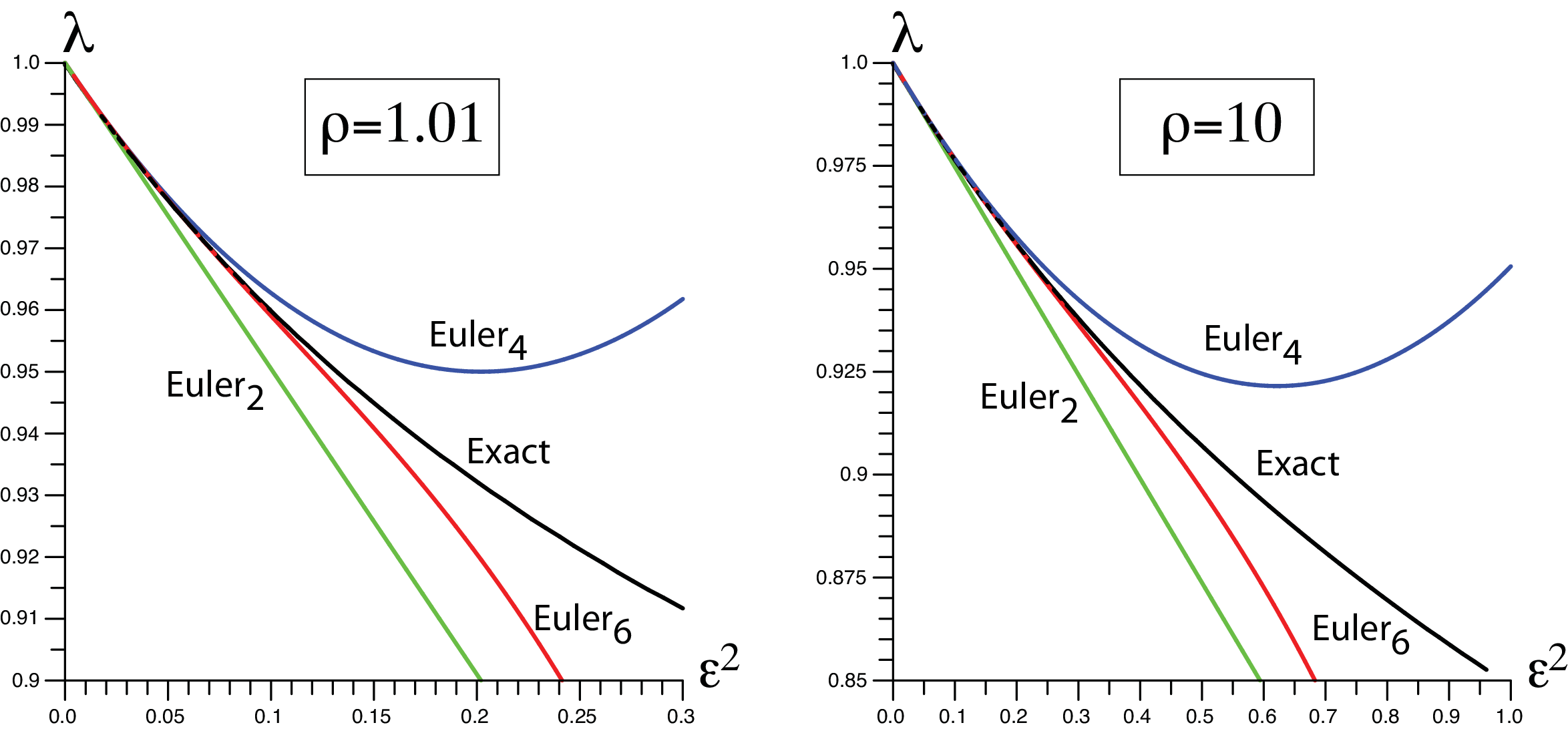}
 \caption{Comparison of the different Euler formulae obtained by expanding the exact solution to order 2 (classical Euler buckling formula), Euler$_4$ and Euler$_6$ for a neo-Hookean potential $C_1=1,C_2=0$. For comparison purpose, we show the critical stretch for mode $n=1$ versus $\epsilon^2$ in which case the graph becomes linear in the limit $\epsilon\to 0$.  Left: $\rho=b/a=B/A=1.01$, Right:  $\rho=b/a=B/A=10$.}
  \label{fig4}
\end{figure}

We are now in a position to look at the asymmetric buckling mode ($n=1$) corresponding to Euler buckling in the limit $\lambda\to 1$. The asymptotic form of Euler criterion cannot be obtained for a general strain-energy density. This is why we choose the Mooney-Rivlin potential, which for $\lambda$ close to 1, corresponds to the most general form of third-order incompressible elasticity (see next Section),
\begin{equation}\label{mooney}
W = C_1(I_1-3)/2 + C_2(I_2-3)/2,
\end{equation}
where $C_1 \ge 0$ and $C_2>0$ are material constants, $I_1=\lambda_1^2 + \lambda_2^2 +\lambda_3^2$, and $I_2 = \lambda_1^2\lambda_2^2 + \lambda_2^2\lambda_3^2 + \lambda_3^2\lambda_1^2$.  
Close to $\lambda=1$, we introduce a small parameter related to the stubbiness ratio
\begin{equation}
\epsilon=k b=\pi b/l,
\end{equation}
and look for the critical buckling stretch $\lambda$ as a function of $\epsilon$ of order $2M$:
\begin{equation}
\lambda=\lambda(\epsilon)=1+\sum_{m=1}^M \lambda_{2m} \epsilon^{2m}+\text{O}(\epsilon^{2M+2}).
\end{equation}
Similarly, we expand $d(\lambda)=\det\vec{Q}(b,a)$ in powers of $\epsilon$, 
\begin{equation}
d(\lambda)=\sum_{m=1}^{M_d} d_{2m} \epsilon^{2m}+\text{O}(\epsilon^{2M_d+2}),
\end{equation}
and solve each order $d_{2m}=0$ for the coefficients $\lambda_{2m}$.  This is a rather cumbersome computation. 
The first non-identically vanishing coefficient appears at order 24 and a computation to order 28 is necessary to compute the correct expression for $\lambda$, which is found to be to order 6 in $\epsilon$
\begin{equation}\label{eul}
\lambda=1+\lambda_{(2)} \epsilon^2 + \lambda_{(4)} \epsilon^4 + \lambda_{(6)} \epsilon^6 + \text{O}(\epsilon^{8}),
\end{equation}
with
\begin{eqnarray}
\label{l2}&& \lambda_{(2)}=- \dfrac{\rho^2+1}{4\rho^2},\\
\label{l4}&& \lambda_{(4)}= \dfrac{ \left( 19\,{C_2}+28\,{C_1} \right) {\rho}^{4}+2\, \left( 53\,{C_2}+62\,{C_1} \right) {\rho}^{2}+19\,{C_2}+28\,{C_1}}{144 (C_1+C_2)\rho^4},
\\
&& \lambda_{(6)} = - \dfrac{1}{4608\,{\rho}^{6} ( \rho^4-1) ( {C_1}+{C_2}) } \times\\
\nonumber &&\phantom{ \lambda_6=}\Bigl[\left( 973\,{C_1}+341\,{C_2} \right)( {\rho}^{10}-1)+ \left( 7073\,
{C_1}+3385\,{C_2} \right) {\rho}^{2}(\rho^6-1)+\\ 
\nonumber&&\phantom{ \lambda_6=9}4392 \ln(\rho)(C_1+C_2)(\rho^6+\rho^4)+4(377 C_2+1141 C_1)(\rho^6-\rho^4)\Bigr],
\end{eqnarray}
where $\rho \equiv B/A = b/a$. 
It is of interest to compare the different approximations. We recover Euler formula by keeping only the term up to $\epsilon^2$ which we denote by Euler$_2$. We define similarly Euler$_4$ and Euler$_6$ by keeping terms up to order 4 and  6 in $\epsilon$. In Fig.\ref{fig4}, we show the different approximations as a function of $\epsilon^2$ for $\rho=1.01$ (on the left) and for $\rho=10$ (on the right). The classical Euler formula is well-recovered in the limit $\epsilon\to 0$ but the Euler$_4$ and Euler$_6$ approximations clearly improve the classical formula for larger values of $\epsilon$. It also appears from the analysis of Euler$_4$ that for $C_2\geq 0$ the classical Euler formula always underestimates the critical stretch for instability.
%


\section{Nonlinear Euler buckling for third-order elasticity}


The analytical result presented in the previous section was formulated in terms of parameters and quantities natural for the computation and the theory of nonlinear elasticity. In order to relate this result to the classical form of Euler buckling, we need to express Eq.~(\ref{eul}) in terms of the initial geometric values $A,B,L$, the axial load acting on the cylinder, and the elastic parameters entering in the theory of  linear elasticity. 

We first consider the geometric parameters. We wish to express the critical load as a function of the initial stubbiness $\nu=B/L$ and tube relative thickness $\rho=B/A$. Recalling that $\epsilon=\pi b/l$ and that $\lambda=l/L, b=\lambda^{-1/2} B$, we have
\begin{equation}\label{epsnu}
\epsilon^2\lambda^3=\pi^2 \nu^2.
\end{equation}
To express $\epsilon$ as a function of $\nu$, we expand $\epsilon$ in powers of $\nu$ to order 6, and solve~(\ref{epsnu}) to obtain\begin{equation}\label{eps}
\epsilon^2=\pi^2\nu^2-3 \pi^4 \lambda_{(2)} \nu^4-(3 \pi^6 \lambda_{(4)}-15 \pi^6 \lambda_{(2)}^2) \nu^6,
\end{equation}
where $\lambda_{(2)}$ and $\lambda_{(4)}$ are defined in~(\ref{l2}-\ref{l4}) and come from the expansion of $\lambda$ in powers of $\epsilon$.

Second, we want to relate the axial compression to the actual axial load $N$. To do so, we integrate the axial stress over the faces of the tubes that is
\begin{equation}
N=-2\pi\int_a^b r \sigma_3 dr.
\end{equation}
Since $\sigma_3$ is constant and given by~(\ref{s3}) we have
\begin{multline}\label{N}
N = - \pi(b^2-a^2) \sigma_3 = - \dfrac{\pi}{\lambda}(B^2-A^2) \sigma_3 \\
  = - \dfrac{\pi}{\lambda^3}(B^2-A^2) \left[{(\lambda^{4}-\lambda)}C_1+{( {\lambda}^{3}-1) } C_2\right].
\end{multline}

Third, we  relate the elastic Mooney parameters $C_1$ and $C_2$ to the classical elastic parameters. Here, we follow Hamilton \textit{et al.} \cite{hailza04,desa05} and write the strain-energy density to third-order for an incompressible elastic material as 
\begin{equation}\label{W3}
W=-2\mu i_2+n_3 i_3,
\end{equation}
where $\mu$ is the usual shear modulus, or second Lam\'e parameter, and $n_3$ is a third-order elasticity constant;
$\mu$ is related to Young's modulus by $E=3\mu$; 
also, in Murnaghan's notation, $n_3 = n$ and in Landau's notation, $n_3 = A$, 
see Norris \cite{norr99} for other notations. 
In \eqref{W3}, $i_1,i_2,i_3$ are the first three principal invariants of the Green-Lagrange strain tensor, related to the first three principal invariants $I_1,I_2,I_3$ of the Cauchy-Green strain tensor by
\begin{equation}
I_1=2i_1+3,\qquad
I_2=4 i_1+4 i_2+3,\qquad
I_3=2 i_1+4 i_2+8 i_3+1,
\end{equation}
Since $I_3=1$, we can solve this linear system for $i_2$ and $i_3$ and write the strain-energy density~(\ref{W3}) as a function of $I_1$ and $I_2$, that is
\begin{equation}
W = \left(\mu + \dfrac{n_3}{8}\right) I_1 
   - \left(\dfrac{\mu}{2} - \dfrac{n_3}{8}\right)I_2,
\end{equation}
which by comparison with~(\ref{mooney}) leads to 
\begin{equation}\label{CC}
C_1 = 2\mu + n_3/ 4, \qquad C_2 = \mu - n_3/ 4.
\end{equation}

To write the nonlinear buckling formula, we consider~(\ref{N}) and first expand $\lambda$ in $\epsilon$ using~(\ref{eul}), then expand $\epsilon$ in $\nu$ using~(\ref{eps}), and, finally, substitute the values of the moduli in terms of the elastic parameters, which yields 
\begin{eqnarray}
&& N = \dfrac{3}{4}{\frac {{\pi}^{3}{B}^{2}\mu\, \left( {\rho}^{4}-1 \right) {\nu}^{
2}}{{\rho}^{4}}}-\\
\nonumber&&\quad \quad {\frac {1}{96}}\,{\frac {{\pi}^{5}{B}^{2} \left( {
\rho}^{2}-1 \right)  \left( 20\,{\rho}^{4}\mu+9\,{\rho}^{4}n_{{3}}+176
\,{\rho}^{2}\mu+18\,{\rho}^{2}n_{{3}}+20\,\mu+9\,n_{{3}} \right) {\nu}
^{4}}{{\rho}^{6}}}+\\
\nonumber&&\quad \quad{{B}^{2}{\pi}^{7}{\nu}^{6}\over 512 \rho^8\mu (\rho^2+1)}\times\\
\nonumber&&\quad \quad  \quad \Bigl[ 323\,{\mu}^{2}{
\rho}^{8}-3\,{n_{{3}}}^{2}-240\,{\rho}^{2}\mu\,n_{{3}}-9\,{\rho}^{2}{n
_{{3}}}^{2}-9\,{\rho}^{10}{\mu}^{2}+9\,{\mu}^{2}+3\,{\rho}^{10}{n_{{3}
}}^{2}+\\
\nonumber&&\quad \quad\quad  1464\,\ln  \left( \rho \right) {\rho}^{6}{\mu}^{2}+1464\,\ln \left( \rho \right) {\rho}^{4}{\mu}^{2}+240\,{\rho}^{8}\mu\,n_{{3}}+
240\,{\rho}^{6}\mu\,n_{{3}}-\\
\nonumber&&\quad \quad \quad 180\,{\rho}^{6}{\mu}^{2}+180\,{\rho}^{4}{
\mu}^{2}+
9\,{\rho}^{8}{n_{{3}}}^{2}+6\,{\rho}^{6}{n_{{3}}}^{2}-6\,{
\rho}^{4}{n_{{3}}}^{2}-323\,{\rho}^{2}{\mu}^{2}-240\,{\rho}^{4}\mu\,n_
{{3}} \Bigr]. 
\end{eqnarray}
While it is not surprising, it is comforting to recover to order $\nu^2$ the classical Euler buckling formula~(\ref{euler}) (using $\rho=B/A$, $\nu=B/L$, and $\mu = E/3$).


\section{Dimensional transition}

\begin{figure}[!ht]
 \centering 
\includegraphics[width=0.6\textwidth]{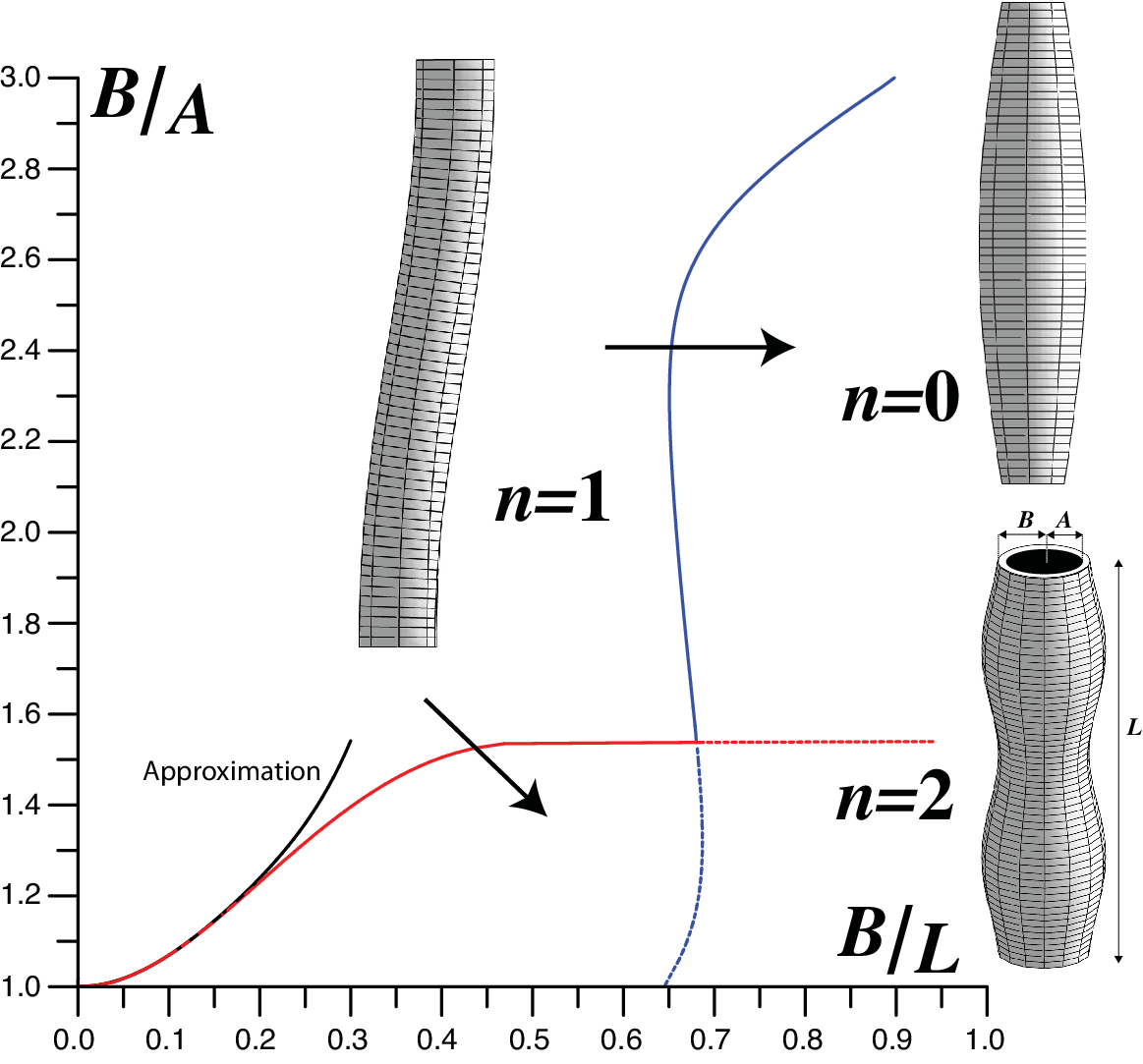}
 \caption{Dimensional transition for a neo-Hookean cylindrical tube of initial length $L$ and initial radii $A$ and $B$. All tubes in the $n=1$ regions will become unstable by buckling. As the tubes get stubbier or thinner (arrows), it will not buckle but instead will be subject to a barreling instability. Note that only the transition curves from mode $n=0$ are shown. Tubes in the barreling regions may be subject to other unstable modes.}
  \label{fig5}
\end{figure}

Finally, we use the buckling formula to compute the transition between modes as parameters are varied. That is, to identify both the geometric values and the axial strain for which there is a transition between buckling and barreling modes. Here we restrict again our attention to the neo-Hookean case (with $C_1=1$). From Fig. \ref{fig2}  and Fig. \ref{fig3}, it appears clearly that for $\epsilon$ small enough there is a transition (depending on the value of $\rho$) from either mode $n=1$ to mode $n=0$ (large $\rho$), or from mode $n=1$ to mode $n=2$ ($\rho$ close to 1) as $\epsilon$ increases. We refer to this transition as a \textit{dimensional transition}, in the sense that the material mostly behaves as a slender one-dimensional structure when it buckles according to mode $n=0$ and mostly as a two-dimensional structure when it barrels with mode $n=2$. Indeed both modes of instability can be captured by, respectively, a one- or a two-dimensional theory. 
For $\rho$ close to unity, the transition $n=0\to n=1$ occurs for small values of $\epsilon$. Therefore in this regime, we can use the approximation~(\ref{eul}) for the barreling curve and substitute it in the bifurcation condition of mode $n=2$. Expanding again this bifurcation condition in $\epsilon$ as well as $\rho$, one identifies the values $\rho_t$ of $\rho$ and $\lambda_t$ of $\lambda$ at which the transition occurs, as
\begin{equation}
\rho_t=1+{3\over4}\epsilon^2-{53 \over 32} \epsilon^4 +{2393\over 384}\epsilon^6 +\text{O}(\epsilon^8),
\qquad 
\lambda_t=1-{1\over4}\epsilon^2+{13 \over 8} \epsilon^4 -{665 \over 96} \epsilon^6+\text{O}(\epsilon^8).
\end{equation}
In terms of the initial stubbiness $\nu=B/L$, we have
\begin{equation}
\rho_t=1+\frac{3}{4}\,{\pi}^{2}{\nu}^{2}-{\frac {17}{32}}\,{\pi}^{4}{\nu}^{4}+{
\frac {161}{384}}\,{\pi}^{6}{\nu}^{6}+\text{O}\left( {\nu}^{8} \right) 
\end{equation}
This  relationship also provides a domain of validity for the Euler buckling formula. For sufficiently slender tube ($\nu$ small), the buckling mode disappears when  $\rho>\rho_t$ at the expense of the $n=2$ barreling mode. For stubbier and fuller tubes, this approximation cannot be used. To understand the dimensional transition, we solve numerically the bifurcation condition, using the adjugate method, for the intersection of two different modes. That is, for a given value of $\rho_*$, we find the value of $\epsilon_*$ such that both the bifurcation for either modes $n=1$ and $n=2$, or modes $n=1$ and $n=0$ are satisfied. If the corresponding  value $\lambda_*$ is the largest value for which a bifurcation takes place, the pair $(\epsilon_*,\rho_*)$ is a transition point. The corresponding transition point in terms of the initial parameters is $(\nu_*={\epsilon_*\over \pi} (\lambda_*)^{3/2}, \rho_*)$. In Fig. \ref{fig5} we show a diagram of all such pairs  for both transitions. 


\section{Conclusion}



This paper establishes a reliable and effective method to study the stability of tubes based on the exact solution of the incremental equations proposed by Wilkes \cite{wi55} within the Stroh formalism.
It then puts the method to use, to  obtain the first geometric and material corrections to Euler buckling. 
The method can be also used to obtain the transition between buckling and barreling modes when a tube becomes unstable. 

The method presented here can be easily generalized to different materials and different boundary conditions. For instance, using the exact solution of the incremental equations proposed in \cite{doha06} for \textit{compressible} materials and the adjugate method, an explicit form of the bifurcation condition in terms of Bessel functions can be obtained by following the steps presented here and various asymptotic behaviors can be obtained. Similarly, a variety of boundary value problems can be analyzed by the adjugate method, such as   the stability problem of a tube under pressure and tension \cite{zhluog08,ha07}, the problem of a tube embedded in an infinite domain \cite{bige01}, and the problem of a tube with coating \cite{ogstha97}. In all these cases, useful asymptotic formulae for the buckling behavior could be obtained by perturbation expansions. 

It is also enticing to consider the possibility of performing an analytical post-buckling analysis of the solutions. Since the solutions of the linearized problem can be solved exactly, a weakly non-linear analysis of the solution should be possible to third-order. This would yield, in principle, an equation for the amplitude of the unstable modes containing much information about the actual amplitude of the unstable modes but also on the localization of unstable modes after bifurcation. We leave this daunting task for another day.


\end{document}